\documentclass[twocolumn,prl,aps,showpacs]{revtex4}
\usepackage{epsfig}
\begin{document}  
\title{
Localization in  the Nonanalytic Quantum Kicked Systems
}
\author{
Jie Liu, Tai-Wang Cheng and  Shi-Gang Chen
}
\affiliation{
Center for Nonlinear Studies,\\
Institute of Applied Physics and
Computational Mathematics, P.O.Box.8009,  100088 Beijing, China
}
\begin{abstract}
Numerical investigations  on nonanalytic quantum kicked systems is presented
in this paper. The  power-law type localization is found
to be universal in the nonanalytic systems just like that
the exponential type in the analytic systems. With increasing
the perturbation strength, a transition from perturbative localization,
to pseudo-integrable regime, 
to dynamical localization and to complete extension
 is clearly
demonstrated.
The different regimes are characterized by
the different relations between the
localization length and the system's  parameters.
 Finally, we investigate the diffusion behavior and find that
  the quantum suppression  is
  relaxed in the nonanalytic systems.
\end{abstract}
\maketitle

\section{introduction}
The quantum dynamics of classically chaotic Hamiltonian systems has
attracted great interest in recent years. Much work has been devoted to
study of the periodically "kicked" systems. The prime examples are
the kicked rotor (KR)\cite{casati1,casati2}
and the kicked hydrogen atom \cite{three}. In these
models, an interesting phenomenon is the dynamical localization,
namely, the quantum suppression of classical diffusion. This phenomenon
was first discovered numerically by Casati {\it et al} \cite{casati1}
 in the
KR model , and later on confirmed by several experiments such as
Rydberg atom in microwave field \cite {four}.
The deviation from maximal chaos in quantum model is entirely related
quantum interference effects only. It was shown that the mechanism of
such a "quantum suppression of classical chaos" is the localization of
all eigenfunctions in the unperturbed momentum space, which similar to the
 well-known Anderson localization in the solid state physics.

 However,
these discussions are mainly restricted to the analytical
systems.
On the other hand,
nonanalytic systems also emerge from the study of more
concrete physical models, such as,
Fermi-acceleration model\cite{fcm}, billiad system \cite{bs1,bs2} and a kicked
particle in 1D infinite square potential well\cite{blgh}.
Recently, some pioneering works have been done towards
 the systems in
which the analyticity condition is broken down \cite{bbb1,bbb2,bbb3}.
 Many novel properties  are found, such as the important role of the
cantori structure in the localization. Their discussions are mainly towards
 a special discontinuous model, where the derivative of the potential
is discontinuous.

The main purpose of this paper is to observe the quantum behavior,
 especially the localization phenomenon in  the
  {\it general  nonanalytic systems}.
 As will be seen later, the nonanalytic systems reveals distinctive
 properties. The power-law type localization is found to be
 universal, just like  the exponential localization in analytic models.
 In particular, the
  power exponent  does not relate to  the
  derivative order of the potential. It rests on the
  convergence rate of the corresponding Fourier series.
   The localization length properly defined in light of the information
 entropy reveals four different regimes with increasing the
 perturbation strength: perturbative localization;
 pseudo-integrable regime; dynamical localization; complete extension.
Each regime is characterized by different relation between the localization
length and the system's parameters.
Moreover, we investigate the diffusion behavior
 and show that the suppression of the  energy diffusion can be
 relaxed in the nonanalytic systems.

The paper is organized as follows.
In Sec.II we introduce the nonanalytic
 Hamiltonian considered in the paper and discuss
 its classical diffusion property.
 The power-law type eigenstates is shown in Sec.III.
  In Sec.IV, we calculate the localization length
   defined in terms of the information entropy. 
   The energy diffusion in nonanalytic systems
   is discussed in Sec.V.
   Finally, Sec.VI is the concluding remarks.
\section{Nonanalytic Models and Classical Dynamics}
The considered Hamiltonian  takes the form (in dimensionless units),
\begin{equation}
\hat H = \frac{\hat p ^2}{2} + k\hat V(q)\delta_T(t),
\,\delta_T(t) = \sum_{m=-\infty}^{+\infty} \delta(t-mT),
\end{equation}
where the potential $V(q)$ is $2\pi$-periodic function, taking the following
form under the Fourier expansion,
\begin{equation}
V(q) = \sum_{n=1}^{\infty}a_n\sin(nq)+b_n\cos(nq).
\end{equation}
To simplify our discussions,  let $a_n =\frac{1}{n^{\alpha}}$, $b_n=0$, then
\begin{equation}
V(q)=V_{\alpha}(q) = C_{\alpha}\sum_{n=1}^{\infty}\frac{\sin(nq)}{n^{\alpha}},
\end{equation}
where the $C_{\alpha}$ is the normalizing constant so that the maximum value
of the potential be unit.
The parameter $\alpha$ represents the convergence rate of Fourier series.
As $\alpha$ goes to infinity, only
the first term of the series remains, which is just the sinusoidal function.
In this case the system is equivalence to the KR model  under a displacement
transformation.
The potential  functions are plotted in Fig.1.
%%%%%%%%%%%%%%%%%%%%%%%%%%%%%%%%%%%%%%%%%%%%%%%%%%%%%
\begin{figure}[!htb]
\begin{center}
\resizebox *{8cm}{8cm}{\includegraphics*{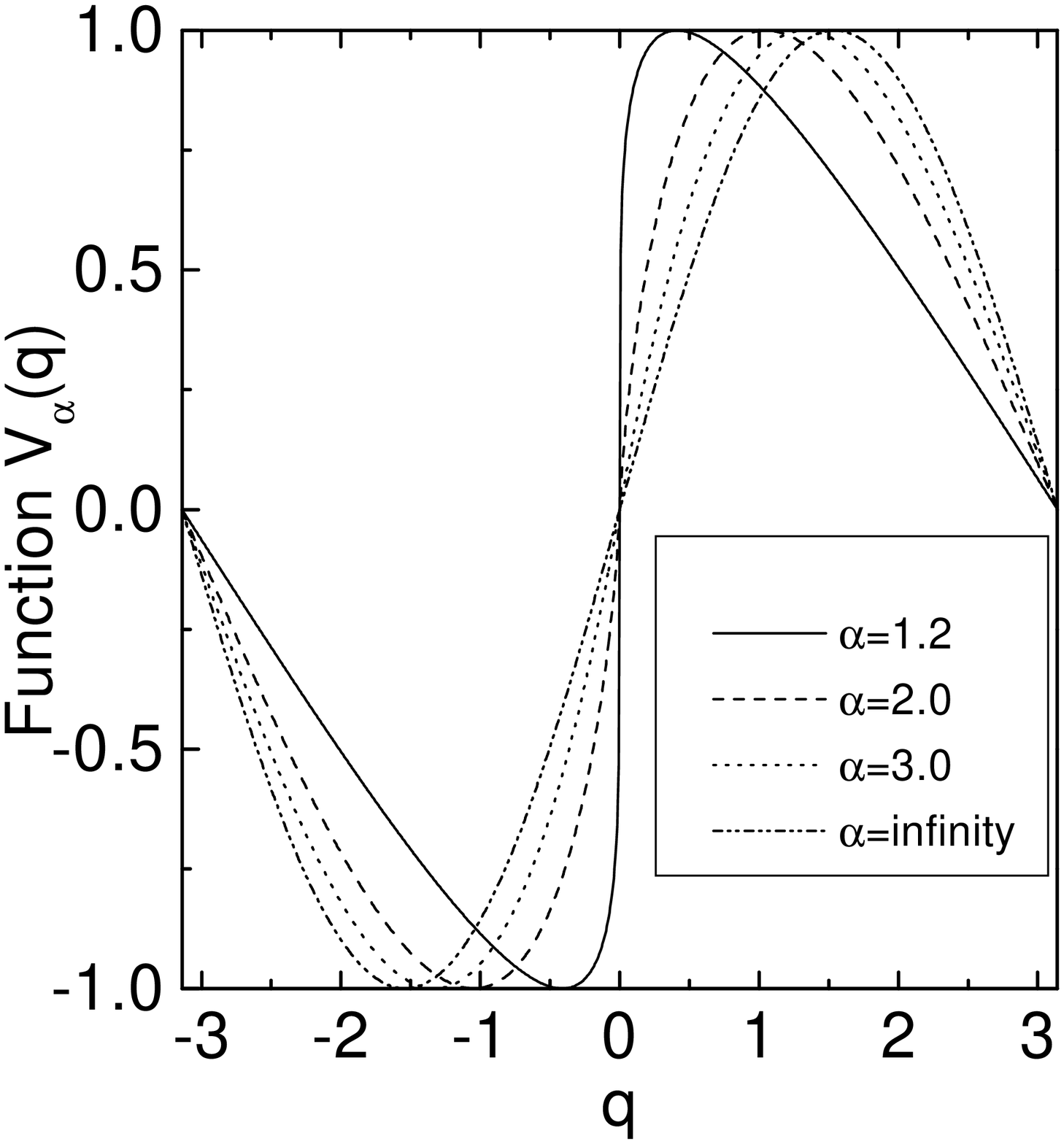}}
\end{center}
\caption{
The  nonanalytic potential functions $V_\alpha (q)$
}
\label{fig:fig1}
\end{figure}
%%%%%%%%%%%%%%%%%%%%%%%%%%%%%%%%%%%%%%%%%%%%%%%%%%%%% 

Before we go further, let us discuss the symmetry problems of the system.
Clearly the time-reversal invariance is kept for any $\alpha$. However, for
$\alpha \ne \infty$, two essential modifications are made: 1) 
$ V_{\alpha}(q)$ turns to be  nonanalytic
($\in C^{(\beta)}$), the order of its
derivability is $\beta =[\alpha]-2$, e.g. its $(\beta+1)$-order
derivative
diverges or discontinuous at $q=0$. Here $[x]$ represents the minimum integer
larger than or equals to
x; 2)
 Spatial symmetry is broken down, the
parity is no longer a good quantum number.
As shown later, these two changes  will  substantially alter the
quantum properties of the system.

Classically, the Hamiltonian (1) corresponds to a twisted map,

$$p_{n+1} = p_{n} + k \frac{\partial {V_{\alpha}(q_n)}}{\partial q_n },$$
\begin{equation}
q_{n+1} = q_{n} + Tp_{n+1} \,\,\, mod (2\pi).
\end{equation}
The classical parameter $K=kT$ determines its classical dynamics.
In our calculations we fix $T=1$.
As $\alpha$ be an integer, the potential can be expressed analyticly,
for example, 
\begin{equation}
\frac{\partial{V_{\alpha}(q)}}{C_\alpha\partial q} =
\left\{
\begin{array}{ll}
ln(2sin(q/2)) & \alpha = 2\\
\frac{1}{12}[(q^2-\pi q) + (2q-\pi)(q-2\pi)] & \alpha =3 , q \in [0,2\pi).\\
\frac{1}{48}[q^2(q-2\pi)^2 -\frac{8}{15}\pi^4] & \alpha =5
\end{array}
\right.
\end{equation}
As well known, the KAM theorem is the footstone of modern dynamics theory. 
It states that invariant tori (KAM tori) of irrational
 winding number will keep up under a small
 perturbation whereas the tori corresponding
  to the rational winding number are broken down.
   Fortunately, the broken tori have  zero-measure
    set in the phase plane. In the original
proof of the theorem, the analyticity is strictly required.
 However, it is found
later on that, the analyticity condition can be relaxed
 to $C^{(3+\epsilon)}$  \cite{herm}
 for a twisted map. Applying
  this  result into our system, it means
  that for $\alpha  > 6$, KAM theorem is available;
   As $\alpha \leq 6$, the system is non-KAM system.

%%%%%%%%%%%%%%%%%%%%%%%%%%%%%%%%%%%%%%%%%%%%%%%%%%%%%
\begin{figure}[!htb]
\begin{center}
\resizebox *{8cm}{8cm}{\includegraphics*{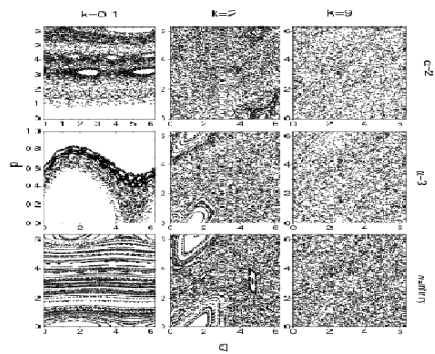}}
\end{center}
\caption{
Local phase plane for various parameters. As $\alpha=2,3$ and $k=0.1$,
the phase plane is ploted by tracing one trajectory originated from
the stochastic region for ten thousands or more periods.
}
\label{fig:fig2}
\end{figure}
%%%%%%%%%%%%%%%%%%%%%%%%%%%%%%%%%%%%%%%%%%%%%%%%%%%%% 
The main characteristic of the non-KAM system is
the broken-up of the KAM tori under a small perturbation
so that the diffusion can take place.
In Fig.2, we plot the local phase planes
 of systems corresponding to  $\alpha = 2,3,\infty$.
  It is shown  that phase plane become more and
   more chaotic as we increase the perturbation strength $K$.
    At small value of $K=0.1$, the phase plane of the nonanalytic systems
    show
     complicated structures, such as  the stochastic webs
      and cantori. The chaotic region is connected and the
       diffusion can occur.
       As $K$ is larger than a critical value $K_{cr}$, the phase planes
       become  fully chaotic, or ergodic. The critical values can be
        estimated approximatly from the losing stability
         of the primary resonances.  Approximately, it gives
$K_{cr}(\alpha=2) \simeq 3.5, K_{cr}(\alpha=3)
\simeq 4.5, K_{cr}(\alpha=5) \simeq 4.6$
and $K_{cr}(\alpha = \infty) \simeq 5.2.$
In the intermediate perturbation $K=2$, even though  the KAM tori
are completely broken down, there are some invariant  structures,
 such as
small islands, cantori.

       In calculating the diffusion
        coefficient for a given $K$, we
 take  1000 points starting from stochastic regions,
  and all the initial trajectories evolve for several thousand periods.
   Averages are taken over 1000 trajectories for each time period.
    It is found that the energy diffusion is
asympotically linear for all values of $K$.
The diffusion coefficient $D (\equiv <\Delta P_n^2>/n) $
(n is the time in unit of T) versus $K$ is
plotted in Fig.3. As the $K$ is large enough i.e. $K > K_{cr}$,
 the diffusion coefficients show
  $D \simeq b_{\alpha} K^2$ in the quasilinear approximation,
   where $b_\alpha =0.843, 0.550 $ and $0.500$ for
   $\alpha$ be $2,3$ and $\infty$, respectively.
    In this case, the phase plane is fully chaotic.
Below these threshold values, diffusion
 coefficients show various behavior for different $\alpha$.
  At $\alpha =2$, a scaling law like $D \sim K^{5/2}$
  is obviously demonstrated. At $\alpha =3$, as $K <0.5$,
   $D \sim  K^2$ and then a $D \sim   K^{7/2}$ scaling law.
At $\alpha =\infty$, as $K > 1$
the diffusion coefficient abruptly rises
  to the quasi-linear approximation regime.
  %%%%%%%%%%%%%%%%%%%%%%%%%%%%%%%%%%%%%%%%%%%%%%%%%%%%%
\begin{figure}[!htb]
\begin{center}
\resizebox *{8cm}{8cm}{\includegraphics*{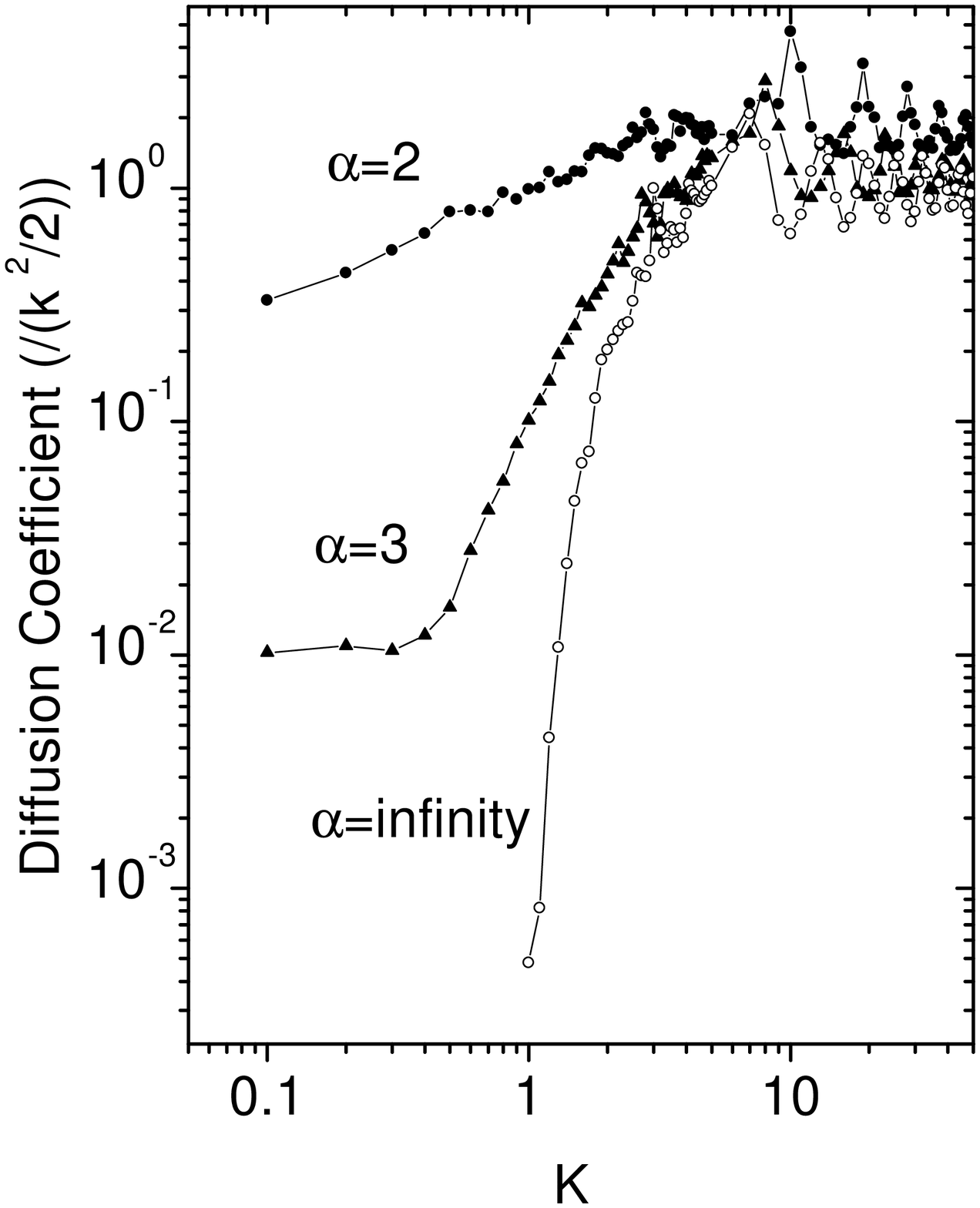}}
\end{center}
\caption{
Classical diffusion coefficients versus the perturbation
 strength for various $\alpha$.
}
\label{fig:fig3}
\end{figure}
%%%%%%%%%%%%%%%%%%%%%%%%%%%%%%%%%%%%%%%%%%%%%%%%%%%%% 

\section{Eigenstates}
Now we turn to the quantum properties of the nonanalytic models
 and to see how the classical chaos manifests itself in the quantum
 dynamics.  The
evolution operator over the period $T$ of the
kick is given by
\begin{equation}
\hat U(T) = exp(-\frac{i\hat P^2 T}{4\hbar})
             exp(-\frac{ik\hat V_{\alpha}(q)}{\hbar})
             exp(-\frac{i\hat P^2 T}{4\hbar}).
\end{equation}
 It is unitary and
satisfies following eigenvalue
equation, $\hat U(T)|\Phi_{\lambda}> =
 e^{-\frac{i\lambda}{\hbar}}|\Phi_{\lambda}> $; Here the eigenphase
 $\lambda$ is real, $\lambda/T$ is so called quasi-energy,
 $\Phi_{\lambda}$ is the eigenstate or Floquet state.

 The Floquet states and the quasi-energy can be obtained by diagonalizing
 $\hat U(T)$ within a large number of plane wave bases
 $|n>, (n=-\frac N 2, ...\frac N 2)$.
In our calculations $N$ is kept at 1024 and $\hbar = T =1$.  The elements of 
matrix  are $U_{nm} = \langle n|\hat U(T)|m \rangle
$.
The fast Fourier transformation (FFT) 
is employed to transform the wave function between the position
representation and momentum representation in our numerical calculations. 
Detailed description of the numerical method refer to \cite{blgh,HLLZ}.

%%%%%%%%%%%%%%%%%%%%%%%%%%%%%%%%%%%%%%%%%%%%%%%%%%%%%
\begin{figure}[!htb]
\begin{center}
\resizebox *{8cm}{8cm}{\includegraphics*{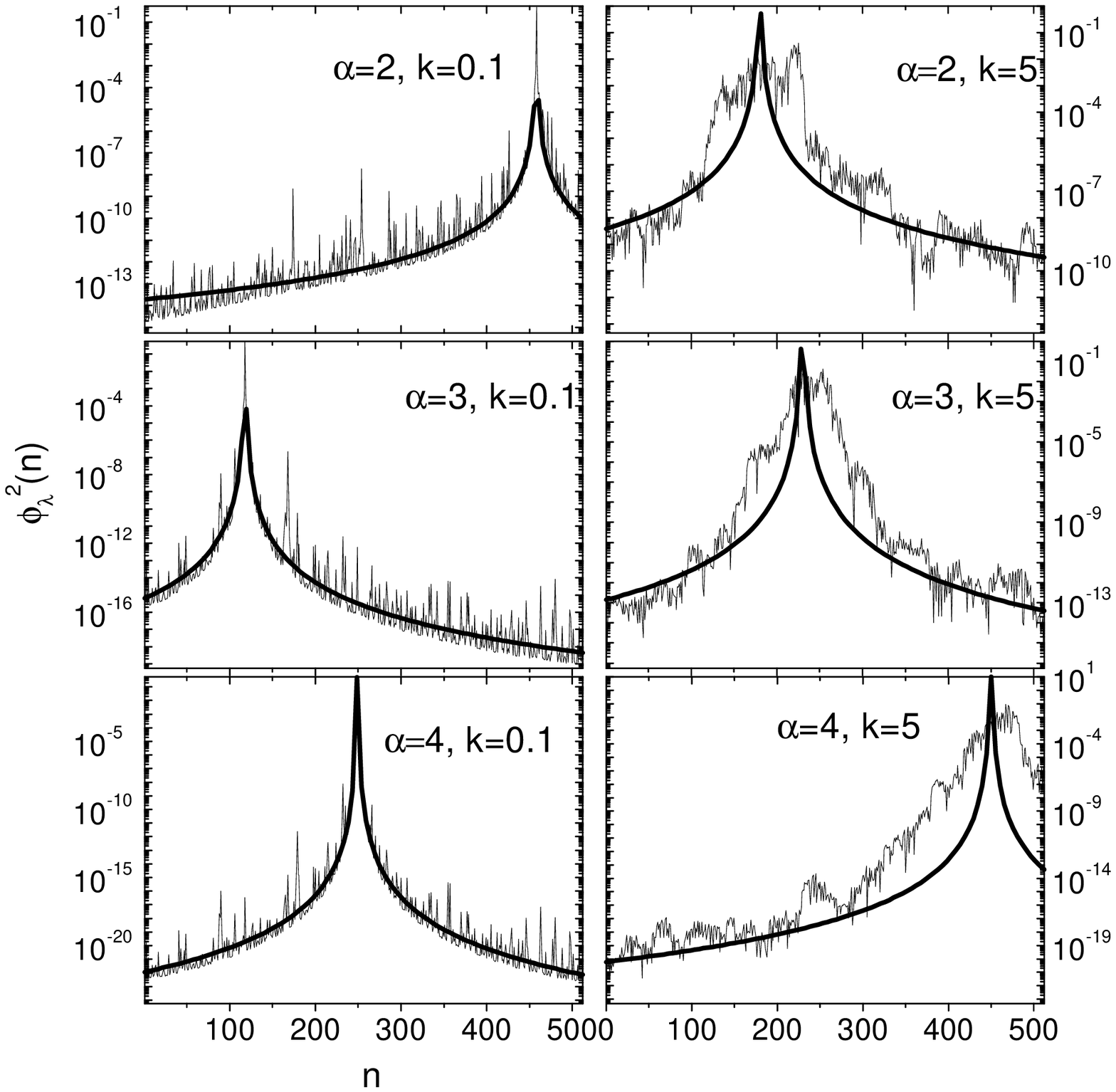}}
\end{center}
\caption{
 The typical eigenvectors in momentum
 representation for different parameters.
The smooth curves represent the Lorentz-like distributions (Eq.8).
The relation between power exponent and the parameter $\alpha$ is
$\gamma = 2\alpha$.        
}
\label{fig:fig4}
\end{figure}
%%%%%%%%%%%%%%%%%%%%%%%%%%%%%%%%%%%%%%%%%%%%%%%%%%%%% 
  In the KR model, the eigenstates are
 always exponentially localized in the momentum space.
 For the nonanalytic systems, the things are quite different.
The time-reversal invariance imposes certain condition on the eigenvector
$\Phi_{\lambda}$, that is,
\begin{equation}
\Phi_{\lambda}(n) = e^{ia}\Phi_{\lambda}^*(-n),
\end{equation}
with $a$ being a constant phase, independent of $n$. So in Fig.4,
 only the half number of the components in the momentum coordinates
  are plotted.
 We find that, for a small perturbation (k=0.1), i.e. in the
 perturbative  regime, the envelope of Floquet states in
 momentum representation can be well fitted by Lorentz-like
 function (see left column of Fig.4):
 \begin{equation}
 |\Phi_{\lambda}(n)|^2 \simeq \frac{c_1}{1 + c_2|n - n_c|^\gamma}.
 \end{equation}
 with $c_1, c_2$ are the fitted constants.

 The power-law localization is clearly
  demonstrated. In particular, the
  power exponent $\gamma$ does not relate to the $\beta$, the
  derivative order of the potential. It rests on the
  convergence rate of the corresponding Fourier series, i.e. a
   simple relation is given numerically,
   \begin{equation}
   \gamma = 2\alpha .
   \end{equation}

   With increasing the perturbation strength, the eigenstate shows a
   chaotic band whose width is proportional to $k$. Outside the band,
   the tails of eigenvector is found to decay, on average, as  power-law.
    The relation between
   the power exponent and the index $\alpha$ expressed by
   Eq.(9) still holds (see right column of Fig.4).

In fact, the power-law type localization of
 the eigenstates can be traced back
to the structure of the matrix $U$.  In the KR model, the values of matrix
elements $U_{mm+n}$ decay faster than exponential when $n$ exceeds the
band width $b$ which is proportional to $k$, thus the elements outside this
diagonal band can be safely regarded as zero.
 Within the band of width $b$, the elements
are proved to be pseudorandom\cite{casati2}.  However, in our model the
situation is different. Careful analysis yields that the elements
outside the band decay as a power law with $|U_{m,m+n}| \sim 1/n^\alpha$. 
We have calculated $\langle U^2\rangle_n ( \equiv \langle
U^2_{m,m+n}\rangle$) (the average is done over $m$). 
The typical slope of the curves over a large range is approximately
minus $2\alpha$ (see Fig.5). And the diagonal band width $b$
 in our model is also found to be
approximately proportional to  $k$ (Fig.5b).
This kind of band random matrix, describes a new class of physical system,
e.g. nonanalytic system, has attracted recent attention\cite{xx}.
%%%%%%%%%%%%%%%%%%%%%%%%%%%%%%%%%%%%%%%%%%%%%%%%%%%%%
\begin{figure}[!htb]
\begin{center}
\resizebox *{8cm}{8cm}{\includegraphics*{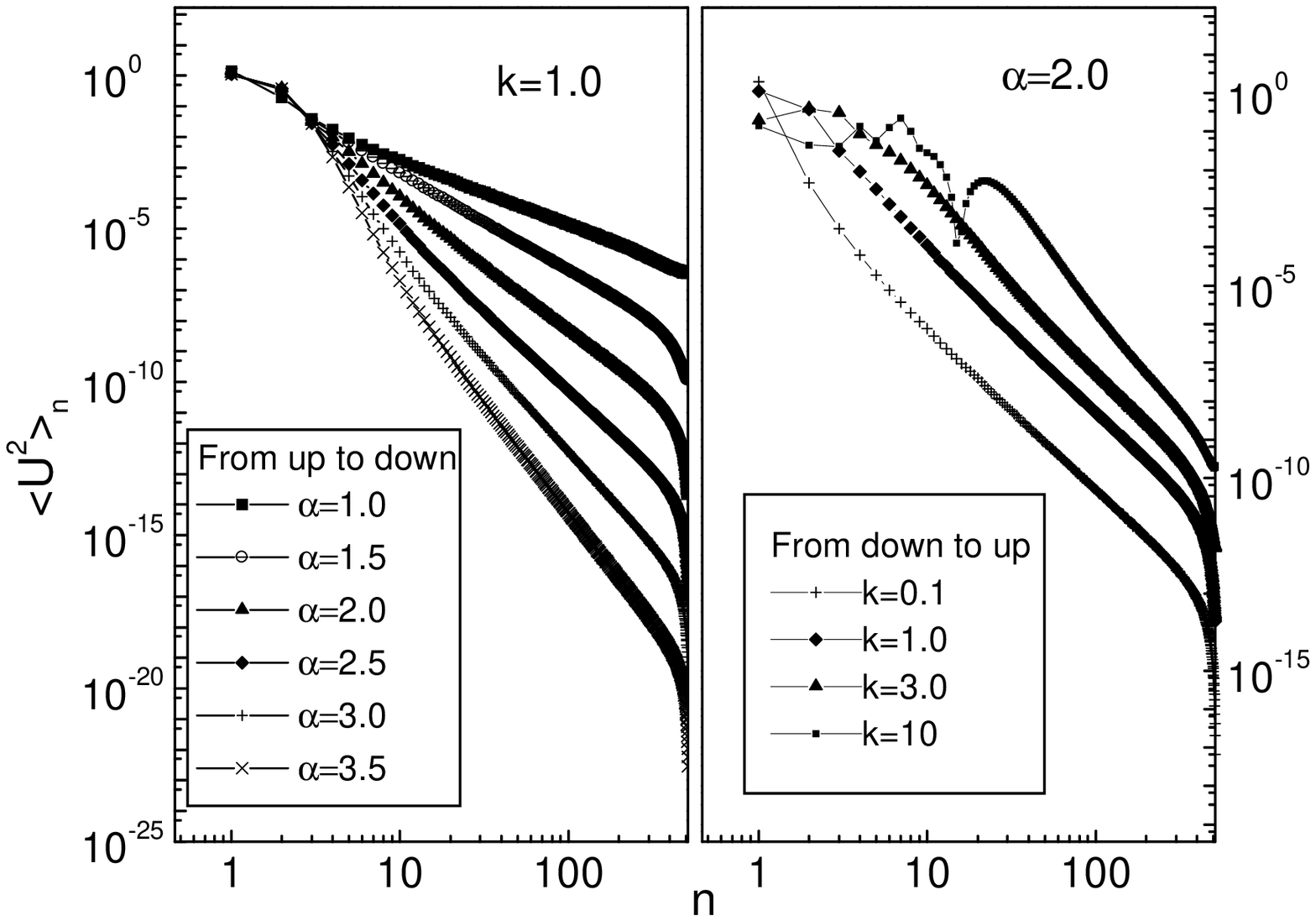}}
\end{center}
\caption{
The averaged matrix elements $<U^2>$ for different parameters
$\alpha$  and $k$. a) right; b) left.
}
\label{fig:fig5}
\end{figure}
%%%%%%%%%%%%%%%%%%%%%%%%%%%%%%%%%%%%%%%%%%%%%%%%%%%%% 

\section{Localization length}
 As shown above, for nonanalytic system, the quantum localization
is  characterized by the power-law decay of the eigenvector in momentum
coordinates. To give a quantitative description of the localization
phenomenon in nonanalytic systems, let us investigate the
localization length  advocated in\cite{iz,casati2}:
\begin{equation}
l = N exp(<\sl{H}> -\sl{H}_R),
\end{equation}
Here $H$ denotes the information entropy for a (N+1)-dimensional eigenvector
$\Phi_{\lambda}(n) (n=-\frac N 2, ...\frac N 2)$,
%\begin{equation}
$\sl{H} \equiv - \sum_{n} \Phi_{\lambda}^2(n) ln(\Phi_{\lambda}^2(n))$,
%\end{equation}
and  the symbol $<...>$ represents an average over all eigenvector;
 The
normalizing coefficient $H_R$ determines the maximal value of entropy.

The above definition is one of commonly used in the
analytic models such as KR model,
it can be extended to nonanalytic systems for
 the information entropy defined for a eigenvector is convergence
 for $\alpha > 0.5$.

 At $\alpha = \infty$, our model is equivalence to the well
 studied model KR model.
 In this case, there exist a additional integral, the
 parity. Together with the time-reversal symmetry, the eigenfunction can
 be expressed by a real function with the relation $\Phi_{\lambda}(n) =
\pm  \Phi_{\lambda}(-n)$. So the total number of the independent
 components
 of   is $N_1 = N/2$, it satisfies following Gaussian distribution
 (e.g. refer  to  \cite{iz}),
 \begin{equation}
 W (|\Phi_{\lambda}(n)|) = 2\sqrt{N_1/\pi}
  e^{-\Phi_{\lambda}^2(n) N_1}.
 \end{equation}
 Then
 this analysis gives a implicit expression of the maxium entropy as
 $H_R \simeq ln(0.48N)$.

 In the case $\alpha \ne\infty$, the parity symmetry is broken,
  the eigenfunctions
 have different real and imaginary parts. Both of them 
    satisfy the Gaussian distribution
  except for $N_1 =N$, then the module of
  the components should satisfy
   following Wigner-type distribution,
   \begin{equation}
     W (|\Phi_\lambda (n)|) =2N_1 |\Phi_\lambda(n)|
     e^{-\Phi_\lambda^2(n) N_1}.
 \end{equation}
Our numerical calculations verify the above distribuation (Fig.6) and
 gives the maximal value of information  entropy as
 $H_R \simeq ln(0.7N)$.
%%%%%%%%%%%%%%%%%%%%%%%%%%%%%%%%%%%%%%%%%%%%%%%%%%%%%
\begin{figure}[!htb]
\begin{center}
\resizebox *{8cm}{8cm}{\includegraphics*{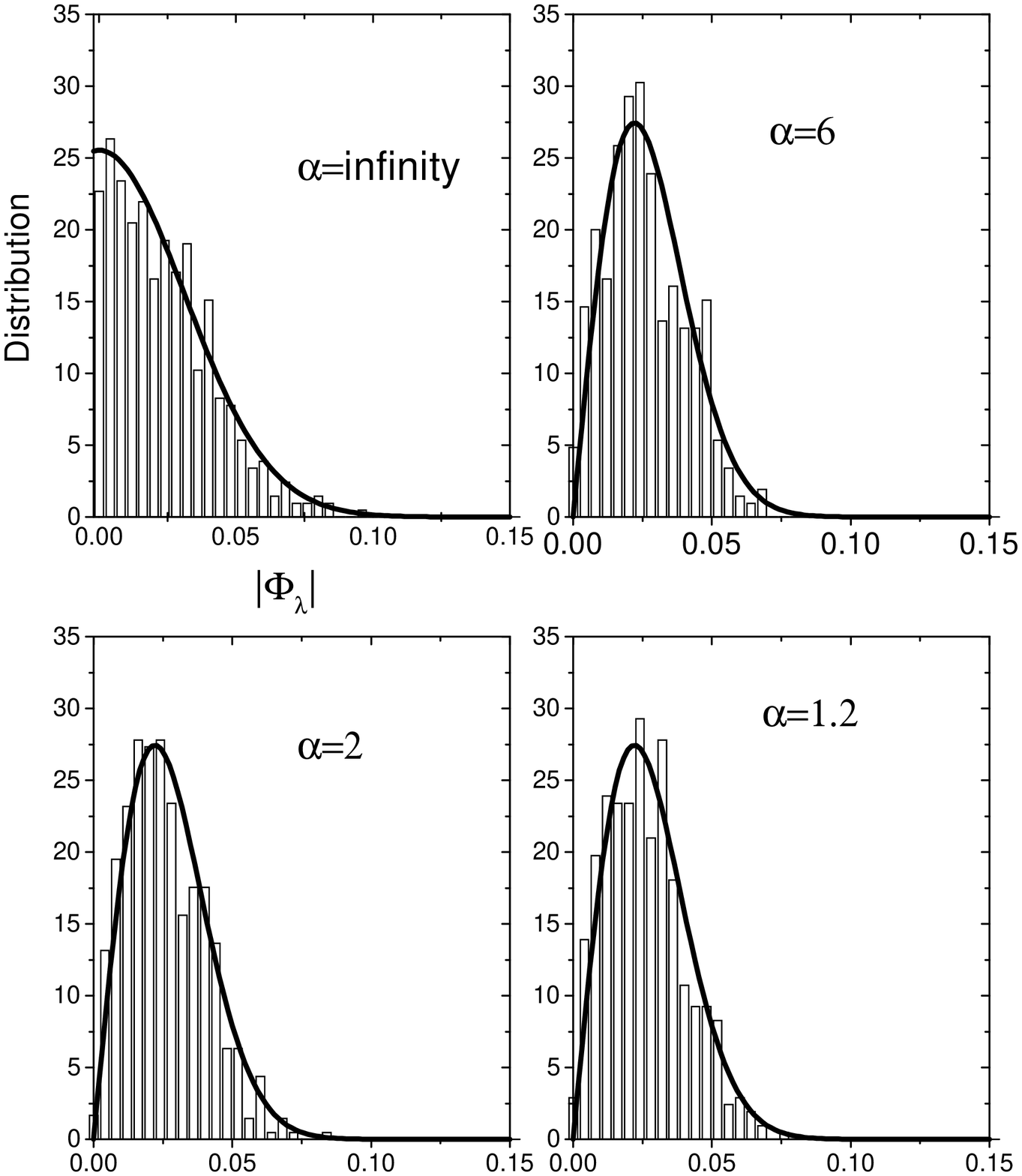}}
\end{center}
\caption{
The distribution of the typical completely extended
 eigenstates for various $\alpha$. The smooth curves are
 distribution expressed by Eq.(11) and Eq.(12), respectively. 
}
\label{fig:fig6}
\end{figure}
%%%%%%%%%%%%%%%%%%%%%%%%%%%%%%%%%%%%%%%%%%%%%%%%%%%%% 

  We   calculate the dependence of the localization length on 
  the perturbation strength  for the different   parameter $\alpha$ (Fig.7a).
 It is shown, the plots of the localization length vs. perturbation amplitude k
 can be divided into four  parts: perturbative localization regime,
 pseudo-integral regime, dynamical  localization regime, and
  saturation regime.
 Each regime is characterized by the different relation between the
 localization length and the perturbation strength.

 In the perturbative regime ($k < 1$),
 average entropy shows linearly proportional  to $k$, i.e. $ln(l) \sim k$
 (Fig.7c).
 In this regime, the Floquet state in momentum representation can be well
 fitted by the Lorentz-like distribution. 
%%%%%%%%%%%%%%%%%%%%%%%%%%%%%%%%%%%%%%%%%%%%%%%%%%%%%
\begin{figure}[!htb]
\begin{center}
\resizebox *{8cm}{8cm}{\includegraphics*{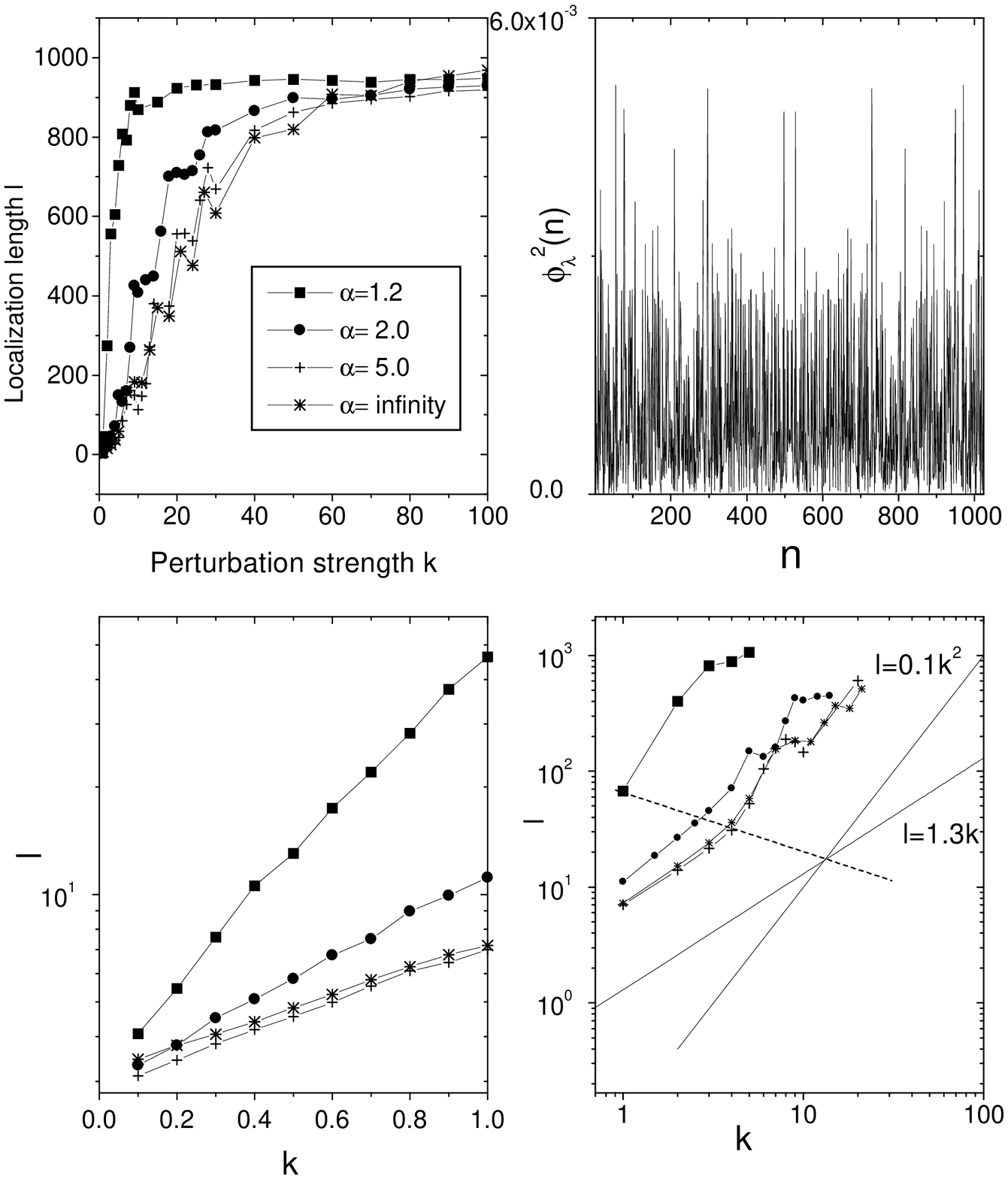}}
\end{center}
\caption{
$\frac{a|b}{c|d}$.
 a) Localization length vs. perturbation strength for different $\alpha$;
      b) In the saturation regime, a typical completely extended
      eigenvector for $\alpha = 5$ and $k=300$;
      c) In the perturbative regime, it is clearly shown $ln(l) \sim k$;
      d) The dashed  line corresponds to the critical values $K_{cl}$ for
      different $\alpha$, approximately.
      In the downside of the dashed line, i.e. pseudo-integral regime,
      it is clearly shown $l \sim k$. In the dynamical
      localization  regime (upside of the dashed line),
       it is clearly shown $l \sim k^2$.        
}
\label{fig:fig7}
\end{figure}
%%%%%%%%%%%%%%%%%%%%%%%%%%%%%%%%%%%%%%%%%%%%%%%%%%%%% 
 If $k$  is large enough, one observe a kind of saturation, i.e.
 the localization
 length almost keep constant, no longer increases with the $k$. This implies
 that eigenstates have become  completely delocalized (Fig.7b).
 The limit value of the length 
  almost tends  to $N$,
  the maximum localization length. 

 In the intermediate $k$ i.e. $N > l$ and $k >1$,
 the
 typical property of the eigenfunctions is the power-law tail.
 The power exponent tightly relates the convergence rate
  of the Fourier  expansion
 of the external potential.
 In this regime, as the perturbation is large enough, i.e. $k > k_{cr}$, the
 phase space will show ergodicity. In this case, it is clearly shown that
 the localization length
 is proportional to the square of the perturbation strength
 i.e. classical diffusion rate $D_{cl}$
 (Fig.7d, upside of the dashed line),
 \begin{equation}
  l \sim k^2.
 \end{equation}
 However, in the case of that perturbation is not so strong, the phase
 space seems not ergodic.
 In this case, even though the KAM tori are broken down,
  the cantori and other invariant structures
 may  act as barriers for the quantum motions,
 if the flux through cantori is less than one
 Planck's cell (for example, see middle column of
 the Fig.2). The eigenstates are expected
 to localized  on these invariant structures.
 Indeed, in such situation the quantum system looks as if classical
 integrable, i.e.
  so called pseudo-intergrable regime
  \cite{bbb1,bbb2,bbb3}. In this case,
  the eigenstates still can  be approximately fitted by
  the Lorentz-like distribution (Fig.8).  Furthermore,
   we  find that in this regime the localization
 length show linearly proportional to the perturbation strength
 (Fig.7d, downside of the dashed line),
 \begin{equation}
  l \sim k.
 \end{equation}
This regime become more and more narrow as
 we decrease the $\alpha$. As $\alpha < 1.5$,
  the classical coefficient tends to infinite which means
  that flow rather than diffusion occurs. In this case, the
  localization length  transits directly from perturbative
  regime to the square scaling-law regime
  (see Fig.7d, case $\alpha = 1.2$).
%%%%%%%%%%%%%%%%%%%%%%%%%%%%%%%%%%%%%%%%%%%%%%%%%%%%%
\begin{figure}[!htb]
\begin{center}
\resizebox *{8cm}{8cm}{\includegraphics*{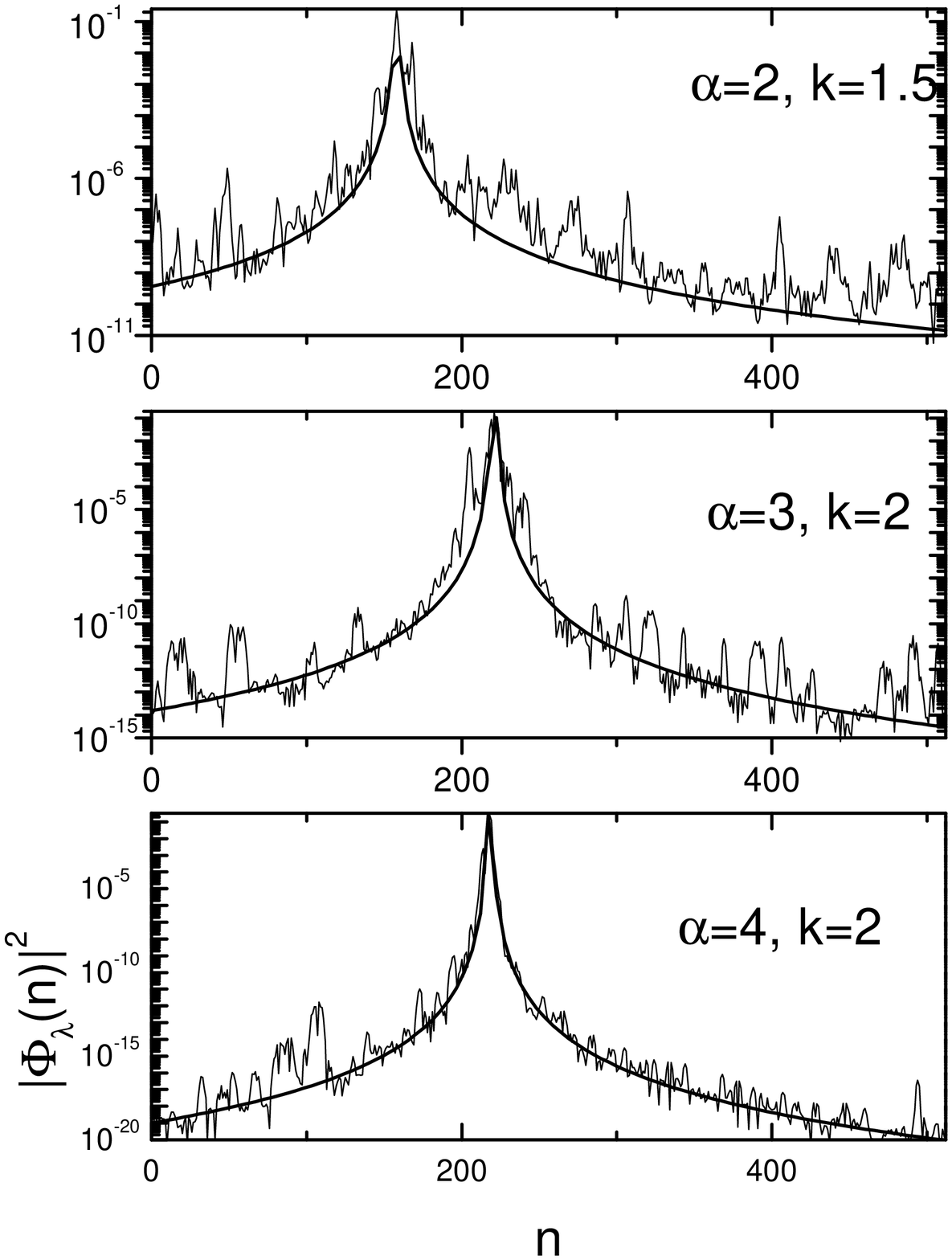}}
\end{center}
\caption{
Typical eigenstates in the pseudo-integral regime.
 The smooth curves represent the Lorentz-like distributions (Eq.8). 
}
\label{fig:fig8}
\end{figure}
%%%%%%%%%%%%%%%%%%%%%%%%%%%%%%%%%%%%%%%%%%%%%%%%%%%%%  
\section{Energy Diffusion}
 The energy diffusion of KR model has been  well studied. After a
typical time scale, the energy
diffusion will be  completely suppressed.
For this model, Floquet states show the strong
localization (exponential decay)   in momentum representation,
 which implies  that,  only a finite
number of Floquet states can  be excited for any initial state. The question
is that what happens to the energy diffusion
in the nonanalytic models, where the Floquet states
 are not so 'strong' localized, only the power-law type localization
 is observed from  the above sections.
In the dynamical localization regime,
out of a chaotic band, the tails of 
the Floquet states behave as
$|\Phi_\lambda (n)|^2 \sim \frac{1}{n^{2\alpha}}$,
 so the average 
energy approximates to
$\sum_n \frac{1}{n^{2\alpha}} \times 2n^2$, which is
divergence for $\alpha \leq 1.5$. From this analysis, we conjecture that
the diffusion suppression will be relaxed as we decrease the parameter
$\alpha$, i.e. lowering the derivable  degree of the system.

We make a  numerical simulation to test the above deduction.
To ensure the precision of the numerical simulation,
both coordinate and momentum spaces should be large enough, so a large number
(N=32768) of Fourier components are used in our computations.
We rescale the coefficients $C_{\alpha}$ so that the classical diffusion
rate $D_{cl} \simeq D_{ql} = k^2/2$ in the quasilinear approximation.
 To satisfy the condition,
$C_{\alpha} = 0.129, 0.219, 0.759,$ and $1.00$ at
$\alpha = 1.52, 1.60, 2.00$ and $\infty$, respectively.
Our results are shown in Fig.9.
The complete suppression of the diffusion is relaxed,
instead, a kind of quantum diffusion
is clearly evident as $\alpha$ be  $1.52$ and $1.60$.
 Our system is isolated from the environment, the observed
 delocalization is completely  due to the nonanalyticity of the the system,
 which leads to the possibility of a  long-range hopping in the momentum
 space. On the other aspect, we find that,
 as $\alpha \geq 2$, the  diffusion is completely suppressed.

 %%%%%%%%%%%%%%%%%%%%%%%%%%%%%%%%%%%%%%%%%%%%%%%%%%%%%
\begin{figure}[!htb]
\begin{center}
\resizebox *{8cm}{8cm}{\includegraphics*{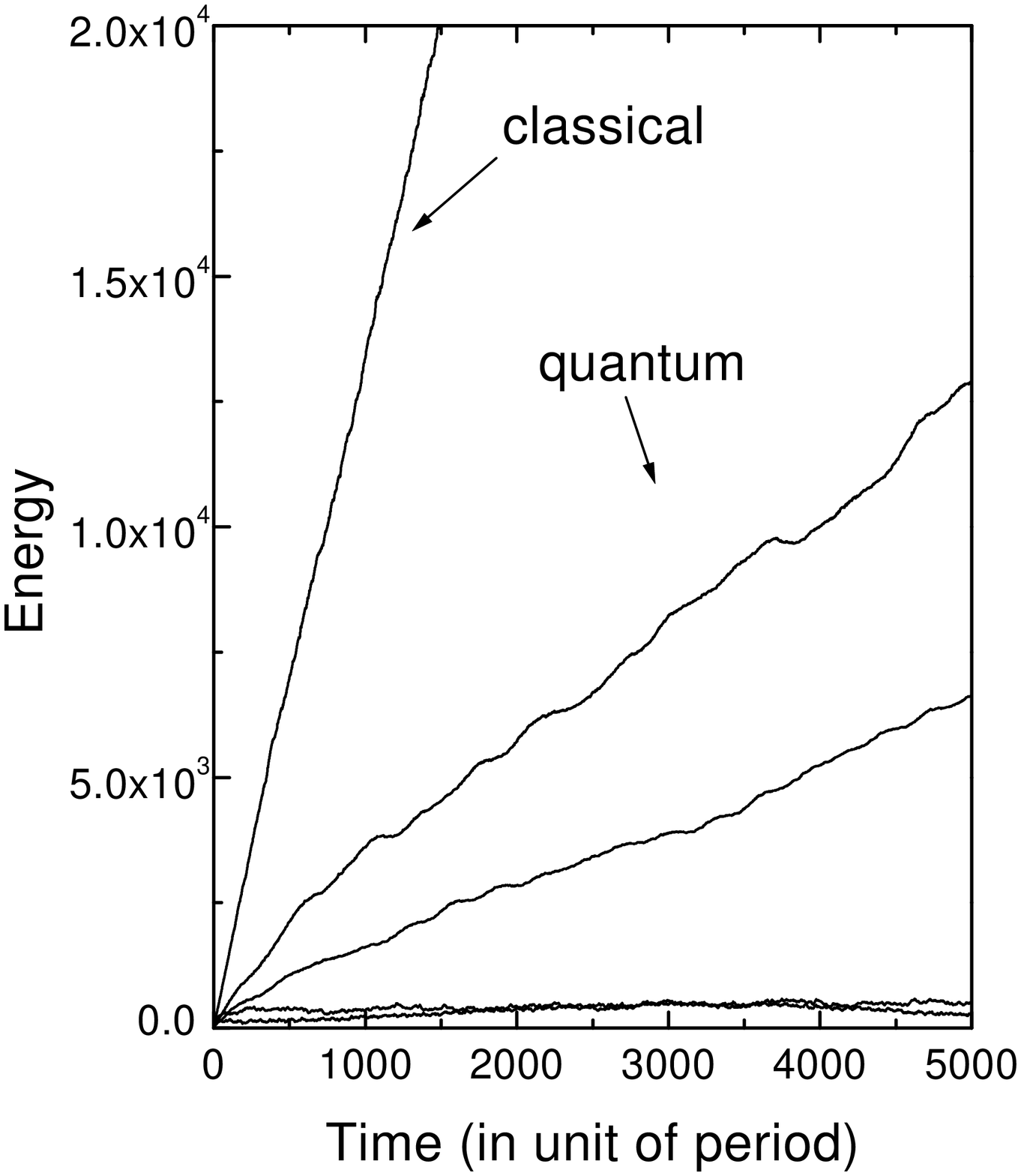}}
\end{center}
\caption{
 The energy diffusion for different parameters $\alpha$
 at $k=6$.
 Classically, the 1000 initial points
  are distributed in $p=0$ with the random $q$ and evolved by
  the map Eq.(4) at  $\alpha =\infty$. For other $\alpha$, since
  the $C_{\alpha}$ is recaled, in the quasilinear approximation,
   the classical evolution shows the similar behavior.
  Quantumly, the initial state is chosen as the
  ground state $|0\rangle$ of
 unperturbed system. The curves from the up to down  correspond to
 $\alpha = 1.52, 1.60, 2.00$ and $\infty$.  It is clearly shown
 that as $ \alpha \geq 2$ the  diffusion is completely suppressed.
 Otherwise, a kind of quantum diffusion occurs. 
}
\label{fig:fig9}
\end{figure}
%%%%%%%%%%%%%%%%%%%%%%%%%%%%%%%%%%%%%%%%%%%%%%%%%%%%%  

\section{Conculsions}
 In summary, nonanalytic systems show many interesting properties,
 the parameter $\alpha$ which represents the convergence rate of the
 Fourier series determines the type of the evolution matrix, therefore
 plays an important role.  Even though our discussion is towards
 a special choice of the potential, the conclusions are generic.
 The Fourier coefficients of any nonanalytic potential take form of
 $c_n/n^{\alpha}$.
  Previously known nonanalytic models are
Fermi-acceleration model\cite{fcm},
billiad system \cite{bs1,bs2} and a kicked
particle in 1D infinite square potential well\cite{blgh}. Their
 common feature is the algebraic decay of the evolution matrix elements,
  the corresponding
parameter $\alpha$ equals to one, two and two, respectively.
In the general nonanalytic case, our studies reveal four different
regimes: perturbative, pseudo-integral, dynamical localization and
complete extension. In each regime,
the relation between the localization
 length and the perturbation strength is
 found to follow different scaling law.
 This finding is in agreement with the arguments
in previous works \cite{bbb1,bbb2,bbb3}

On the other hand,  it was found that declocalization
 occurs if one let the system interact with an environment. This became
 an active point attracting much attention theoretical and experimentally
 in recent years.
 Our findings in this paper suggest that declocalization can be realized
 by breaking the analyticity of the system. We hope our works will
 stimulate the studies in the  direction. 

\section*{Acknowledgement}
Liu  gives his
special thanks to Dr.B.Li for many useful suggestions and discussions.
He  also thanks Prof.S.Y.Kim for drawing his attention to nonanalytic
models. This project is supported by National Nature Science Foundation of
China and Climbing Project of Nonlinear Science.

%\begin{thebibliography}

Figures Captions

Fig.1 The  nonanalytic potential functions $V_\alpha (q)$.

Fig.2 Local phase plane for various parameters. As $\alpha=2,3$ and $k=0.1$,
the phase plane is ploted by tracing one trajectory originated from
the stochastic region for ten thousands or more periods.

Fig.3 Classical diffusion coefficients versus the perturbation
 strength for various $\alpha$.

Fig.4  The typical eigenvectors in momentum
 representation for different parameters.
The smooth curves represent the Lorentz-like distributions (Eq.8). 
The relation between power exponent and the parameter $\alpha$ is
$\gamma = 2\alpha$.

Fig.5  The averaged matrix elements $<U^2>$ for different parameters
$\alpha$  and $k$. a) right; b) left.

Fig.6 The distribution of the typical completely extended
 eigenstates for various $\alpha$. The smooth curves are
 distribution expressed by Eq.(11) and Eq.(12), respectively.
Fig.7 $\frac{a|b}{c|d}$.
 a) Localization length vs. perturbation strength for different $\alpha$;
      b) In the saturation regime, a typical completely extended
      eigenvector for $\alpha = 5$ and $k=300$;
      c) In the perturbative regime, it is clearly shown $ln(l) \sim k$;
      d) The dashed  line corresponds to the critical values $K_{cl}$ for
      different $\alpha$, approximately.
      In the downside of the dashed line, i.e. pseudo-integral regime,
      it is clearly shown $l \sim k$. In the dynamical
      localization  regime (upside of the dashed line),
       it is clearly shown $l \sim k^2$.

Fig.8 Typical eigenstates in the pseudo-integral regime.
 The smooth curves represent the Lorentz-like distributions (Eq.8).

Fig.9 The energy diffusion for different parameters $\alpha$
 at $k=6$.
 Classically, the 1000 initial points
  are distributed in $p=0$ with the random $q$ and evolved by
  the map Eq.(4) at  $\alpha =\infty$. For other $\alpha$, since
  the $C_{\alpha}$ is recaled, in the quasilinear approximation,
   the classical evolution shows the similar behavior.
  Quantumly, the initial state is chosen as the
  ground state $|0\rangle$ of 
 unperturbed system. The curves from the up to down  correspond to
 $\alpha = 1.52, 1.60, 2.00$ and $\infty$.  It is clearly shown
 that as $ \alpha \geq 2$ the  diffusion is completely suppressed.
 Otherwise, a kind of quantum diffusion occurs.

%\end{thebibliography}
%\end{multicols}
\end{document}